\documentstyle[prl,aps,preprint]{revtex}
\begin{document}
Physica C {\bf 218} (1993) 197-207\\

\begin{center}
   {\Large \bf Large size two-hole bound states in $t$-$J$ model}
\end{center}
       M. Yu. Kuchiev$^{1,2}$ and O. P. Sushkov$^{1,3}$\\
{\em $^1$ School of Physics, The University of New South Wales\\
   P.O.Box 1, Kensington, NSW 2033, Australia\\
$^2$ A.F.Ioffe Physical-Technical Institute, 194021 St. Petersburg, Russia\\
$^3$ Budker Institute of Nuclear Physics, 630090 Novosibirsk, Russia}

\begin{abstract}
Several series of shallow large-size ($l \gg lattice~ spacing$)
two-hole bound states are found in the two-dimensional $t-J$ model.
Their binding energies depend exponentially on the inverse value of the
hole-spin-wave coupling constant. Their parity is connected with the
angular wave function in a nonstandard way. The possible role of these
states in the formation of superconducting pairing is considered.
\end{abstract}

\vskip 2.cm
\begin{center}
spin fluctuations, energy gap.
\end{center}

\newpage
\section{Introduction}

The problem of mobile holes in the $t$-$J$ model is closely related to
high-$T_c$ superconductivity. The $t$-$J$ model with less than
half-filling is defined by the Hamiltonian
\begin{equation}
\label{1}
  H = H_t + H_J
    = -t \sum_{<nm>\sigma} ( d_{n\sigma}^{\dag} d_{m\sigma} + \mbox{H.c.} )
    + J \sum_{<nm>}  {\bf S}_n{\bf S}_m,
\end{equation}
where $d_{n\sigma}^{\dag}$ is the creation operator of a hole with spin
$\sigma$ ($\sigma= \uparrow, \downarrow$) at site $n$ on a
two-dimensional square lattice. The $d_{n\sigma}^{\dag}$ operator acts
in the Hilbert space where there is no double electron occupancy. The
spin operator is
${\bf S}_n = {1 \over 2} d_{n \alpha}^{\dag}
${\boldmath $\sigma$}$_{\alpha \beta} d_{n \beta}$.
$<nm>$ are neighbour sites on the lattice. Below we set $J=1$.

At half-filling (one hole per site) the $t$-$J$ model is equivalent to
the Heisenberg antiferromagnet model \cite{Hir5,Gro7} which has the long-range
antiferromagnetic order in the ground state \cite{Oit1,Hus8,Manousakis}.
The problem is the behavior of the system under doping by additional holes.

Single particle properties in the $t$-$J$ model are by now well
established. A single hole is an object with complex structure due to
virtual admixture of spin excitations. It has been shown that a single
hole has the ground state with a momentum of ${\bf k}=(\pm \pi/2, \pm
\pi/2)$. The energy is almost degenerate along the line $\cos k_x+\cos
k_y=0$ which is the edge of the magnetic Brillouin zone (see e.g.
Refs.\cite{Tru8,Sch8,Shr8,Kan9,Bon9,Dag0,Mart1,Liu2,Sus1,Sus2}).

The two-hole problem is much more complicated. There is no doubt that
for $t \ll 1$ there are short-range $d$- and $p$-wave two-hole bound
states with a binding energy $\Delta E \approx -0.25$. These bound
states could be relevant to the superconductivity. However the value of
$t$ corresponding to the realistic high-$T_c$ superconductors is $t
\approx 3$ (see, e.g. Refs.\cite{Esk0,Fla1,BCh3}). The early works on
exact diagonalizations for finite-size
clusters\cite{Bon9,Has9,Luk0,DRY0,Ito0,Rie1} indicated the existence of
a short-range two-hole bound state for large $t$. However recent
investigations \cite{Ede2,Bon2,Poilblanc,Cher3} have demonstrated that
the short-range two-hole bound state vanishes at $t \approx 2-3$.
Moreover, in our opinion even for smaller values of $t$ the short-range
Coulomb repulsion in realistic systems should destroy this bound state
(see the discussion in Ref.\cite{Cher3}). These arguments lead us to
the conclusion that theory predicts no short-range bound states at
physical values of $t$. This conclusion makes the investigation of the
long-range dynamics in the $t-J$ model very important.

The long-range dynamics is determined by the interaction of
a hole with spin-wave excitations. Calculation of the hole-spin-wave
coupling constant $f$ is straightforward if one treats $H_t$ as
perturbation and uses the condition $zS \gg 1$ ($z$ is the number of
neighbour sites).
The value of the coupling constant obtained in this way is
$f={1\over 2}zt=2t$ (see e.g.Ref.\cite{Kan9,Mart1,Liu2}). The calculation of
$f$
for realistic $zS$ and arbitrary $t$ has been done in recent
works\cite{Suh3,Suhf}.

In the present work we prove the existence of shallow large-size
two-hole bound states which appear due to the exchange of a spin-wave.
It is demonstrated that there are several infinite series of such
states. Their binding energies depend exponentially on the radial quantum
number: $E_n \propto e^{-\gamma \pi n}$, where $\gamma$ is a
numerical constant, $n=0,1,\ldots$. Very interesting is the relation
between the parity and the angular wave function of these states. A
state with an angular wave function without nodes (in this sense it is
$s$-wave) has negative parity. A state with two nodes in the angular
wave function (in this sence it is $p$-wave) has positive parity. This
property may be considered as an additional negative ``internal''
parity of the pair. The nonstandard parity relation is caused by two
physical reasons. First, the ground state of the system is the
antiferromagnetic one. Second, the bottom of single-hole dispersion is
at the face of the Brillouin zone.

Our paper has the following structure.
In Sec.II we derive an effective Hamiltonian describing the dynamics of
the system in the long-range region. It is presented in terms of dressed holes
and spin-waves. The nontrivial properties of the system are caused by
the strong interaction between holes and spin-waves and the contact
interaction between the holes. This section is based on the results of
Refs.\cite{Cher3,Suh3,Suhf}.
In Sec.III we show that the exchange of a spin-wave between two holes
creates the long-range attracting potential which decreases as $1/r^2$
with the separation $r$ between the holes. This potential generates
several infinitive series of bound states of the two holes. In Sec.IV
we discuss the parity and the symmetry properties of these states. In
Sec.V we compare the analytical solution with the results of exact numerical
diagonalization of the equation for the bound states. In concluding Sec.VI we
discuss scenario of possible superconducting pairing based upon the results of
the present work.

\section{Interaction between the holes with opposite spins.
Effective Hamiltonian for the dressed holes.}

 We treat the $J$ term of the Hamiltonian (\ref{1}) using the
linear-spin-wave approximation (see Ref.\cite{Manousakis} for a review).
Define the following Fourier transformations
\begin{equation}
\label{2}
a_{\bf q}^{\dag} =
  \sqrt{2\over{N}} \sum_{n \in \uparrow}
   S_n^- e^{i{\bf q} \cdot {\bf r}_n},
\hspace{1.cm}
b_{\bf q}^{\dag} =
  \sqrt{2\over{N}} \sum_{n \in \downarrow}
  S_n^+ e^{i{\bf q} \cdot {\bf r}_n},
\end{equation}
where our notation  $n \in \uparrow$ ($n \in \downarrow$) means that
site $n$ is on the up sublattice (down sublattice).
Introducing the Bogoliubov canonical transformation,
\begin{equation}
\label{3}
\alpha_{\bf q}^{\dag} = u_{\bf q} a_{\bf q}^{\dag} -
v_{\bf q} b_{\bf -q}, \hspace{1.cm}
\beta_{\bf q}^{\dag} = u_{\bf q} b_{\bf q}^{\dag} -
v_{\bf q} a_{\bf -q},
\end{equation}
we write the Heisenberg Hamiltonian $H_J$ as
\begin{equation}
\label{4}
H_J=E_0 +\sum_{\bf q}\omega_{\bf q}(\alpha_{\bf q}^{\dag}\alpha_{\bf q}
+\beta_{\bf q}^{\dag}\beta_{\bf q}),
\end{equation}
where $E_0$ is the antiferromagnetic background energy.
The summation over ${\bf q}$ is restricted inside the Brillouin
zone of one sublattice where $\gamma_{\bf q}= {1\over 2} (\cos q_x+\cos q_y)\ge
0$.
The spin-wave dispersion and the transformation coefficients are given by
\begin{eqnarray}
\label{5}
\omega_{\bf q}&=&2\sqrt{1-\gamma_{\bf q}^2} \to
\sqrt{2}|{\bf q}|, \ at \ q \ll 1 \nonumber \\
u_{\bf q} &=& \sqrt{{1\over{\omega_{\bf q}}}+{1\over{2}}},\\
v_{\bf q} &=& -sgn(\gamma_{\bf q})
\sqrt{{1\over{\omega_{\bf q}}}-{1\over{2}}}.\nonumber
\end{eqnarray}
The spin waves $\alpha_{\bf q}$ and $\beta_{\bf q}$
have definite values of the spin projections. Due to Eqs.(\ref{2}),(\ref{3})
$\alpha_{\bf q}^{\dag} |0\rangle$ has $S_z=-1$, and
$\beta_{\bf q}^{\dag} |0\rangle$ has $S_z=+1$. Here $|0\rangle$ is the
quantum N\'{e}el state wave function.

 A single hole on the antiferromagnetic background is a composite object due to
strong virtual admixture of the spin excitations. Therefore its
dispersion is sufficiently complicated. Fortunately it may be well
approximated by the simple analytical formula which was found in
Refs.\cite{Sus2,Suhf}.
\begin{eqnarray}
\label{6}
\epsilon_{\bf k}&=&\sqrt{\Delta^2/4+4t^2(1+y)}-
\sqrt{\Delta^2/4+4t^2(1+y)-4t^2(x+y)\gamma_{\bf k}^2}
+{1\over4}\beta_2(\cos k_x -\cos k_y)^2\\
 \Delta &\approx& 1.33,\ x \approx 0.56, \ y \approx 0.14 \nonumber
\end{eqnarray}
The parameters $\Delta,x,y$ are some combinations of the ground
state spin correlators\cite{Sus2}. The $\beta_2$ term gives the
dispersion along the edge of the Brillouin zone. It is rather weak and
may be estimated as $\beta_2 \sim 0.1\times t$ at
$t \ge \Delta/4$. Near the band bottom ${\bf k}_0=(\pm \pi/2, \pm \pi/2)$
the dispersion (\ref{6}) can be presented in a usual quadratic form
\begin{equation}
\label{7}
\epsilon_{\bf k}\approx{1\over2}\beta_1 k_1^2+
{1\over2}\beta_2 k_2^2, \hspace{1.5cm}\beta_2 \ll \beta_1,
\end{equation}
where $k_2$ is projection of the momentum along the face of magnetic Brillouin
zone, $k_1$ is orthogonal projection of the momentum, see Fig.1. Due to
Eq.(\ref{6})
\begin{equation}
\label{8}
\beta_1 \approx \left\{ \begin{array}{ll}
 4{{x+y}\over{\Delta}}t^2 \approx 2.1t^2 & \mbox{for $t\ll \Delta/4=0.33$},\\
 {{x+y}\over{\sqrt{1+y}}}t\approx 0.65t  & \mbox{for $t\gg \Delta/4$}.
                        \end{array}
                \right.
\end{equation}
A comparison with numerical calculations (see e.g.
Refs\cite{Kan9,Dag0,Mart1,Liu2,Sus2,Suhf}) shows that simple
formulas (\ref{6}),(\ref{8}) underestimate $\beta_1$ by 10-20\%. This
produces no serious effect on the results of the present work, but
certainly this correction can to be included.

 The wave function of a single hole may be written in the form
$\psi_{{\bf k}\sigma}=h_{{\bf k}\sigma}^{\dag} |0\rangle$.
$h_{{\bf k}\sigma}^{\dag}$ is the creation operator of a dressed hole
\begin{eqnarray}
\label{9}
h_{{\bf k}\uparrow}^{\dag} &=&
\sqrt{2\over{N}} \sum_{n \in \downarrow}
h_{n \uparrow}^{\dag} e^{i{\bf k} \cdot {\bf r}_n}, \\
h_{{\bf k}\downarrow}^{\dag} &=&
  \sqrt{2\over{N}} \sum_{n \in \uparrow}
h_{n \downarrow}^{\dag} e^{i{\bf k} \cdot {\bf r}_n}, \nonumber
\end{eqnarray}
It was demonstrated in Refs.\cite{Sus2,Suhf} that $h_{n \sigma}^{\dag}$
may be well approximated by the following simple expressions:
\begin{eqnarray}
\label{10}
&&h_{n \uparrow}^{\dag}\approx \nu d_{n\uparrow}^{\dag} + \mu S_n^+
\sum_{\delta}
 d_{n+{\delta} \downarrow}^{\dag},\nonumber\\
&& \nu = {1\over 2} \biggl[ {{3/2+2S_t}\over S_t} \biggr]^{1/2}, \hspace{1.cm}
   \mu = {t \over {[S_t(3/2+2S_t)]^{1/2}} },\\
&& S_t = \sqrt{9/16+4t^2}, \nonumber
\end{eqnarray}
where {\boldmath $\delta$} is a unit vector corresponding to one step
in the lattice. This form of trial wave function leads to the dispersion
(\ref{6}).

 The effective interaction of a composite hole with a long wave-length
spin-wave is of the form (see, e.g. Refs.\cite{Kan9,Mart1,Liu2,Suh3,Suhf})
\begin{eqnarray}
\label{11}
&&H_{h,sw}= \sum_{{\bf k},{\bf q}}g({\bf k},{\bf q})
\biggl(h_{{\bf k}+{\bf q}\downarrow}^{\dag}
h_{{\bf k}\uparrow} \alpha_{\bf q}+h_{{\bf k}+{\bf q}\uparrow}^{\dag}
h_{{\bf k}\downarrow} \beta_{\bf q} + H.c. \biggr),\\
&&g({\bf k},{\bf q})=2f\sqrt{{2\over N}}(\gamma_{\bf k}u_{\bf q}+
\gamma_{{\bf k}+{\bf q}}v_{\bf q}).\nonumber
\end{eqnarray}
In the perturbation theory limit ($t \ll \Delta/4$) and at $zS \gg 1$
($z$ is the number of neighbour sites) the coupling constant
is $f \approx f_{0}=2t$ (see e.g. Ref.\cite{Kan9,Mart1,Liu2}). Account of the
first $1/zS$ correction gives $f \approx f_1={8\over 3}t$, and
summation of all $1/zS$ series gives $f=3.4t$ for $t \ll \Delta/4$
\cite{Suh3}. For an arbitrary $t$ the coupling constant was calculated
in Ref.\cite{Suhf}. The plot of $f$ as a function of $t$ is presented
in Fig.2. For small t: $f \approx 3.4t$. For large $t$ the coupling
constant is $t$-independent: $f \approx 2$.

  The interaction between the two holes can be caused by the exchange
of single (or several) soft spin-wave. Alongside with that there is a
contact hole-hole interaction. One can say that it is due to the
exchange of several hard spin-wave excitations. Now we are going to
derive the Hamiltonian of the contact hole-hole interaction.
Interaction between two holes at large momentum transfer ($q \sim 1$)
has been obtained in Ref.\cite{Cher3} using a variational approach and
$1/zS$ expansion.
\begin{eqnarray}
\label{12}
&&\Gamma({\bf k}_3\uparrow,{\bf k}_4\downarrow;{\bf k}_1\uparrow,{\bf k}_2
\downarrow)
 \approx {8\over N}\biggl[ A\gamma_{{\bf k_1}-{\bf k_3}}
 + B(\gamma_1 \gamma_3 + \gamma_2 \gamma_4)
 + {C\over 2} (\gamma_{{\bf k}_1+{\bf k}_3}+\gamma_{{\bf k}_2+{\bf k}_4})
    \biggr] \delta_{12,34},\\
&&A = 16t\nu \mu^3(1-7\mu^2) - {1\over4}-2\mu^2-18.5\mu^4+84\mu^6 +
10\alpha t \nu^3 \mu^3, \nonumber\\
&&B = 8t\nu \mu(1-9\mu^2+32\mu^4) - 4\mu^2(1-4\mu^2)(1-12\mu^2),\nonumber\\
&&C = {2\over 3}\alpha t \nu \mu^3.\nonumber
\end{eqnarray}
Here $\nu$ and $\mu$ are the parameters of the wave function
(\ref{10}). We denote for simplicity $\gamma_{{\bf k}_i}=\gamma_i$. The
functions $A$ and $C$ in formula (\ref{12}) have been derived in the
first order in $\alpha$, where $\alpha$ is the coefficient in front of
the transverse contribution to the Heisenberg energy: ${\bf S}_n{\bf
S}_m \to S^z_n  S^z_m+
{\alpha \over 2} ( S^+_n S^-_m + S^-_n S^+_m )$. The physical value is
$\alpha=1$. The higher order contributions to $A$ and $C$ are
essential. In order to fit them one has to set $\alpha$ in
Eq.(\ref{12}) somewhere in between $\alpha=0.5$ and $\alpha=1$. The
expression for $B$ in formula (\ref{12}) corresponds to the Ising
background ($\alpha=0$). The difference is that $A(\alpha=0)$
and $C(\alpha=0)$ are small and therefore the corrections are crucial.
The coefficient $B$ is not small and the $\alpha$ correction is not so
important.

Expression (\ref{12}) includes the contributions from the single
spin-wave exchange as well as the many spin-wave exchange. It is
important to separate them because, as it will be clear from further
consideration, they play quite a different role in the physics of
large-size bound states.

 Let us demonstrate now that the $B$-term in (\ref{12}) is
due to the single spin-wave exchange, see Fig.3. For the Ising
background there is no hole dispersion, the frequency of the spin-wave
is equal $\omega_{\bf q}=2$, and the Bogoliubov's parameters in
Eq.(\ref{5}) are $u_{\bf q}=1$, $v_{\bf q}=0$. Therefore due to
Eq.(\ref{11}) the single spin-wave exchange contribution (Fig.3)
is equal
\begin{equation}
\label{13}
-{{8f^2}\over{N}}{1\over{(-\omega_{\bf q})}}
(\gamma_1 \gamma_3 + \gamma_2 \gamma_4)=
{{4f^2}\over{N}}(\gamma_1 \gamma_3 + \gamma_2 \gamma_4).
\end{equation}
We see that the single spin-wave exchange (\ref{13}) possesses the same
kinematic structure as the $B$-term in (\ref{12}). Let us verify that
their absolute values are also identical.
The interaction (\ref{12}) has been derived in first order in $1/zS$
expansion. In this approximation for $t \ll \Delta/4$ the hole
spin-wave coupling constant is $f={8\over 3}t$ (see Ref.\cite{Suh3}).
Substituting this value into expression (\ref{13}) one can easily check
that it perfectly agrees with $B={32\over 9}t^2$ obtained from
Eq.(\ref{12}). At large $t$ the agreement between (\ref{13}) and
$B$-term in Eq.(\ref{12}) cannot be perfect because the latter comes
from an approximate variational solution. Nevertheless numerically the
agreement is good enough. For example for $t=3$ the coupling constant
is $f\approx 1.8$ (see Fig.2) and the value of $B$ obtained from
formula (\ref{13}) is $B\approx 1.6$. From expression (\ref{12}) one
gets $B\approx 2.2$. This consideration demonstrates that the $B$-term
in the short-range interaction
(\ref{12}) accounts for the single spin-wave exchange. Therefore $A$
and $C$ terms describe the contact hole-hole interaction. Due to
formula (\ref{12}) the Hamiltonian of contact interaction is of the form
\begin{equation}
\label{14}
H_{hh}\approx{8\over N} \sum_{1,2,3,4}\biggl[ A\gamma_{{\bf k_1}-{\bf k_3}}
 + {C\over 2} (\gamma_{{\bf k}_1+{\bf k}_3}+\gamma_{{\bf k}_2+{\bf k}_4})
    \biggr]h_{3\uparrow}^{\dag}h_{4\downarrow}^{\dag}
h_{2\downarrow}h_{1\uparrow} \delta_{12,34}.
\end{equation}
We denote for simplicity $h_{{\bf k}_i}=h_i$.

  Let us formulate the result of the present section. Starting
from the Hamiltonian of the t-J model (\ref{1}) which is expressed in
terms of operators $d_{n\sigma}$ we show that the dynamics of holes on
the antiferromagnetic background is described by the effective Hamiltonian
\begin{equation}
\label{15}
H_{eff}=\sum_{\bf k}\epsilon_{\bf k}h_{\bf k}^{\dag}h_{\bf k}+
\sum_{\bf q}\omega_{\bf q}(\alpha_{\bf q}^{\dag}\alpha_{\bf q}
+\beta_{\bf q}^{\dag}\beta_{\bf q}) + H_{h,sw} + H_{hh}
\end{equation}
which is expressed in terms of composite operators $h_{{\bf k}\sigma}$
and spin-waves $\alpha_{{\bf q}}, \beta_{{\bf q}}$. The interactions
$H_{h,sw}$ and $H_{hh}$ are given by Eqs.(\ref{11}) and (\ref{14}).

\section{Two-hole bound states}

Consider a system of two holes with total momentum ${\bf P}=0$ and the
projection of spin $S_z=0$.
Let $g_{{\bf k}}$ be the wave function in the momentum representation
describing the state of the system with the relative momentum  $2{\bf k}$:
\begin{equation}
\label{16}
|{\bf P}=0 \rangle = {{1}\over{\sqrt{N}}} \sum_{\bf k} g_{\bf k}
 h^{\dag}_{{\bf k}\uparrow} h^{\dag}_{- {\bf k} \downarrow} |0\rangle.
\end{equation}
The function $g_{{\bf k}}$
satisfies the equation of the type of Bethe-Salpeter one:
\begin{equation}
\label{17}
 (E-2\epsilon_{\bf k}) g_{\bf k} = \sum_{\bf p}
\Gamma(E;{\bf k},-{\bf k};{\bf p},-{\bf p}) g_{\bf p}.
\end{equation}
Summation is carried out over the magnetic Brillouin zone
($\gamma_{\bf p} \ge 0$). The kernel of this equation can be presented in the
form
\begin{equation}
\label{Gam}
\Gamma=\Gamma_{sw}+\Gamma_{contact}.
\end{equation}
The term $\Gamma_{sw}$ in (\ref{Gam}) is the contribution from the
single spin-wave exchange (see Fig.3) for which we find with the help
of Eq.(\ref{11})
\begin{equation}
\label{Gamsw}
\Gamma_{sw}=
-{{16f^2}\over{N}}
{{(\gamma_{\bf -k}u_{\bf q}+\gamma_{\bf p}v_{\bf q})
(\gamma_{\bf -p}u_{\bf q}+\gamma_{\bf k}v_{\bf q})}
\over{E-\epsilon_{\bf p}-\epsilon_{\bf k}-\omega_{\bf q}}},
\end{equation}
where ${\bf q}={\bf p}+{\bf k}$.
The term $\Gamma_{contact}$ in Eq.(\ref{Gam}) desribes the many
spin-wave exchange. From Eq.(\ref{14}) we derive
\begin{equation}
\label{Gamcon}
\Gamma_{contact}=
{8\over N}\biggl[ A\gamma_{{\bf k}-{\bf p}}+C \gamma_{{\bf k}+{\bf p}} \biggr].
\end{equation}

Let us apply Eq.(\ref{17}) to the problem of the large-size bound
state. In this case one can restrict the consideration to the two-hole
dynamics near the band bottom in one pocket of the Brillouin zone, for
example near the point ${\bf k}_0=(\pi/2,\pi/2)$ (see Fig.1). We have
to allow $k_1$ to be negative as well as positive. The hole state with
negative $k_1$ is outside the Brillouin zone, but it is equivalent to
one with ${\bf k'}={\bf k}-{\bf g}$,
which is inside the Brillouin zone. Here ${\bf g}=(\pi,\pi)$ is the
vector of the inverse magnetic lattice. We will show that the single
spin-wave exchange $\Gamma_{sw}$ (\ref{Gamsw}) plays the crucial role
in the formation of these states. Therefore we consider it first. Later
we will take into account the term $\Gamma_{contact}$ (\ref{Gamcon}).
Note also that the binding energy as well as the kinetic energy of the
pair will be shown to be small compared to the spin-wave energy.
Therefore we can neglect $E-\epsilon_{\bf p}-\epsilon_{\bf k}$ compared
to $\omega_{\bf q}$ in the denominator in Eq.(\ref{Gamsw}). As a result we
obtain
\begin{equation}
\label{Gamma}
\Gamma \approx \Gamma_{sw}\approx {{8f^2}\over{N}}{{q_1^2}\over{q^2}}=
{{8f^2}\over{N}}{{(k_1+p_1)^2}\over{({\bf k}+{\bf p})^2}}.
\end{equation}
We have shifted the zero of momentum to the center of the poket.
Equation (\ref{17}) is reduced with the help of (\ref{Gamma}) to the simple
form
\begin{equation}
\label{BS}
(E-\beta_1 k_1^2-\beta_2 k_2^2) g_{\bf k}
=8f^2\int {{d^2p}\over{(2\pi)^2}}{{(k_1+p_1)^2}\over{({\bf k}+{\bf
p})^2}} g_{\bf p}.
\end{equation}
The last equation is invariant in respect to the transformation ${\bf
k} \to -{\bf k}$, ${\bf p} \to -{\bf p}$. That is why we can classify the wave
function by its behaviour under the transformation:
\begin{equation}
\label{g}
g_{\bf k} \to g_{-{\bf k}}={\cal R}g_{\bf k},
\end{equation}
where we introduce the quantum number ${\cal R}= \pm 1$. Let us stress
that ${\cal R}$ is not the parity of the state because ${\bf k}$ is the
deviation from ${\bf k}_0$ which is not the center of symmetry of the
Brillouin zone. The relation between the quantum number ${\cal R}$ and
the parity will be discussed below. From (\ref{BS}, \ref{g}) we get
\begin{equation}
\label{E}
(E-\beta_1k_1^2-\beta_2k_2^2)g_{\bf k}=8f^2{\cal R} \int{{d^2p}
\over{(2\pi)^2}}{{(k_1-p_1)^2} \over{({\bf k}-{\bf p})^2}}g_{\bf p}.
\end{equation}
Using the Fourier transformation
\begin{equation}
\label{Four}
\psi({\bf r})= \int g_{\bf k} e^{i{\bf k}{\bf r}} {{d^2k} \over{(2\pi)^2}}
\end{equation}
we can rewrite it in the coordinate representation as usual Schrodinger
equation:
\begin{equation}
\label{Sch}
\biggl(-\beta_1{{{\partial}^2}\over{{\partial}x_1^2}}-
\beta_2{{{\partial}^2}\over{{\partial}x_2^2}} +U({\bf r})\biggr)\psi
=E\psi,
\end{equation}
where the potential energy is
\begin{equation}
\label{U}
U({\bf r})={\cal R}(2f^2/ \pi){{x_2^2-x_1^2}\over{r^4}}.
\end{equation}
It is convenient to perform the rescaling $x_1=\sqrt{\beta_1}\eta$,
$x_2=\sqrt{\beta_2}\zeta$ and to
introduce the polar coordinates $\eta=\rho \cos{\varphi}$,
$\zeta=\rho \sin{\varphi}$. In new variables equation (\ref{Sch}) has the form
\begin{equation}
\label{Epsi}
\biggl(-{{1}\over{\rho}}
{{\partial}\over{{\partial}{\rho}}} {\rho}
{{\partial}\over{{\partial}{\rho}}}
-{1\over{{\rho}^2}}{{{\partial}^2}\over{{\partial}{\varphi}^2}}
+{{V(\varphi)}\over{{\rho}^2}}\biggr)\psi=E\psi,
\end{equation}
where
\begin{eqnarray}
\label{V}
&& V({\varphi})={\cal R}Fa{{\sin^2\varphi-a\cos^2\varphi}
\over{(\sin^2\varphi+a\cos^2\varphi)^2}}, \\
&& F=2f^2/(\pi \beta_1), \hspace{1.cm} a=\beta_1/\beta_2. \nonumber
\end{eqnarray}
The plot of the interaction constant $F$ as a function of $t$ is presented
in Fig.2. Typical mass ratio in which we are interested is $a \sim 3 - 10$.
The variables in equation (\ref{Epsi}) are separated, the solution is of the
form
\begin{equation}
\label{psi}
\psi=R(\rho)\Phi(\varphi).
\end{equation}
{}From (\ref{Epsi}), (\ref{psi}) we find the following equations:
\begin{equation}
\label{Phi}
\biggl(-{{d^2}\over{d\varphi^2}}
+V(\varphi)\biggr)\Phi(\varphi)=\lambda \Phi(\varphi),
\end{equation}
\begin{equation}
\label{ER}
\biggl(-{{1}\over{\rho}}
{d\over{d\rho}} {\rho}
{d\over{d\rho}}
+{{\lambda}\over{{\rho}^2}}\biggr)R(\rho)=ER(\rho).
\end{equation}
The boundary condition for $\Phi(\varphi)$ follows from the symmetry
relation (\ref{g}):
\begin{equation}
\label{bound}
\Phi(\varphi+\pi)={\cal R}\Phi(\varphi).
\end{equation}
The boundary condition for $R(\rho)$ we will discuss later.

 The crucial thing is
the sign of the parameter $\lambda$, which is the eigenvalue of the
angular equation (\ref{Phi}). The latter has the form of the usual
Schrodinger equation in which the function $V(\varphi)$ plays the role
of a potential and the eigenvalue $\lambda$ plays the role of an
energy. Note that $V(\varphi)$ (\ref{V}) depends on the symmetry of the
wave function (\ref{bound}) through the symmetry parameter ${\cal R}$.
If one neglects $V(\varphi)$ then the solution of Eq.(\ref{Phi}) is
trivial, $\lambda=m^2 \ge 0,~m=0,1,\ldots$. The point is that
$V(\varphi)$ provides the regions of attraction, where $V(\varphi)<0$.
For ${\cal R}=+1$ these regions are in the vicinity of points
$\varphi=0$ and $\varphi= \pi$, e.g. in the direction perpendicular to
the edge of the Brillouin zone. For ${\cal R}=-1$ they are near points
$\varphi= \pi/2$ and $\varphi= -\pi/2$, e.g. in the direction along the
edge of the Brillouin zone, see Fig.1. Hence one can expect that there
are solutions of Eq.(\ref{Phi}) with negative eigenvalues $\lambda$.
They provide the
attracting potential in the radial equation (\ref{ER}) and, as a
result, the bound states for the two holes. Let us show these solutions
do exist and examine their properties.

Consider first the case when the potential $V(\varphi)$ (\ref{V}) is
small, $F \ll 1$. Then the solutions of Eqs.(\ref{Phi}), (\ref{Phi}) in
zero approximations are $\Phi^{(0)}_m(\varphi)=\exp(im\varphi)$,
$\lambda^{(0)}_m=m^2$, $m=0, \pm1, \ldots.$

For the first level we have $m=0$, $\Phi^{(0)}_0(\varphi)=1$,
$\lambda^{(0)}_0=0$. Using Eq.(\ref{V}) one can easily verify that
first order correction to the eigenvalue vanishes:
$\lambda^{(1)}_0={1\over{2\pi}}\int_0^{2\pi}V(\varphi)d\varphi=0$.
Therefore only the
second order correction, which is definitely negative, is essential:
$\lambda_0\approx\lambda^{(2)}_0<0$. This proves the existence of
negative eigenvalue of Eq.(\ref{Phi}) for small $F$. The considered
solution belongs to ${\cal R}=1$ symmetry.

Consider now the next level. It is degenerate: $m=\pm1,
\Phi^{(0)}_m(\varphi)=\exp(\pm i \varphi), \lambda^{(0)}_m=1$. The
potential $V(\varphi)$ splits it. The splitting appears in the first
order of perturbation theory because the level is degenerate. As a
result one component of the doublet, $\lambda_1$, goes down with increasing $F$
as:
\begin{equation}
\label{lambda}
\lambda_1 \approx 1-{1\over{\pi}}\int_0^{2\pi}V(\varphi)\cos^2\varphi
d\varphi=1-{{2aF} \over{(1+\sqrt{a})^2}}.
\end{equation}
Therefore for sufficiently large $F$ one has to expect that this
eigenvalue is to become negative. The corresponding eigenfunction
possesses ${\cal R}=-1$ symmetry.

For large $F$ both eigenvalues $\lambda_0$ and $\lambda_1$ decrease,
but $\lambda_1$ does it faster. Therefore for some $F$ there is an
intersection of $\lambda_0$
with $\lambda_1$. The reason for this fact is the following. If ${\cal
R}=-1$ then the potential $V(\varphi)$ (\ref{V}) has minima near the
points $\varphi=\pm {{\pi} \over{2}}$ which become deeper with increase
of $F$. As a result the ground-state function $\Phi_{{\cal
R}=-1}(\varphi)$ is strongly localized in the vicinity of these points.
Therefore we can approximate the potential $V(\varphi)$ by the
potential of a harmonic oscillator near the points $\varphi=\pm {{\pi}
\over{2}}$. This gives the following asymptotic
expression for the eigenvalue $\lambda_1$ which is valid if $Fa \gg 1$:
\begin{equation}
\label{lam1}
\lambda_1 \approx -Fa+ \sqrt{Fa(3a-1)}-{{15a^2-15a+2} \over{4(3a-1)}}.
\end{equation}
This proves that $\lambda_1$ becomes negative for sufficiently large $F$.

The corresponding expression for $\lambda_0$ is more complicated,
because the potential $V(\varphi)$ for ${\cal R}=1$ can possess either
two or four minima for different values of $a$. Therefore we present
only the lower estimation for $\lambda_0$:
\begin{equation}
\label{lam0}
\lambda_0 > V_{min}(\varphi)|_{{\cal R}=1}=\left\{ \begin{array}{ll}
-F{{(a+1)^2} \over{8(a-1)}}  & \mbox{for $a \ge 3$} \\
-F & \mbox{for $a \le 3$}.
\end{array}
\right.
\end{equation}
Comparing inequality (\ref{lam0}) with expression (\ref{lam1}) we
really see that $\lambda_1$ goes down with increase of $F$ more rapidly
than $\lambda_0$.

There is another parameter, the mass ratio $a$, which governs the value
of $\lambda$. It is clear that the
higher $a$ the more our two-dimensional system looks like a
one-dimensional one. The influence of the potential in one-dimension is
stronger, the binding energies are deeper. Therefore both $\lambda_0$
and $\lambda_1$ decrease with the increase of $a$. Due to formulas
(\ref{lam1}), (\ref{lam0}) $\lambda_1$ decreases faster than $\lambda_0$.

Fig.4 presents the results of numerical solution of the eigenvalue
problem (\ref{Phi}),(\ref{bound}). In agreement with above
consideration $\lambda_0$  (${\cal R}=+1$) is negative for any positive $F$.
The negative $\lambda_1$ (${\cal R}=-1$) appears when $F$ is large enough.

  For sufficiently large $F$ there will appear other solutions with negative
$\lambda$. However we restrict our consideration to the lowest solutions
because they are most important for the problem of bound states. For
illustration we
present the eigenfunctions $\Phi(\varphi)$ for $F=1.4$ and $a=7$ in
Fig.5. The solution with ${\cal R}=-1$ has two nodes ($\varphi=0,\pi$),
and in this sense it is ``$p$-wave''. There are no nodes in solution
with ${\cal R}=+1$, and in this sense it is ``$s$-wave''.

Consider now the radial equation (\ref{ER}). For negative $\lambda$
there is the long-range attractive potential in Eq.(\ref{ER}). As a
result for every such $\lambda,~\lambda=\lambda_0, \lambda_1<0,$ there
appear the infinite series of bound states for the two holes. We can
find their energies using the semiclassical approximation.
\begin{equation}
\label{En}
E_n \approx -{{4|\lambda|}\over{{\rho}_0^2}}
\exp\biggl(-2-{{2\pi}\over{\sqrt{|\lambda|}}}(n+3/4)\biggr).
\end{equation}
Here $\rho_0$  is a short-range cut off parameter.
The point is that the interaction $-1/{\rho}^2$ is too singular at the
origin and therefore it gives the ``fall'' of a particle to the center.
To stabilize the solution we introduce infinite repulsion at $\rho \le
\rho_0$. It is obvious that the cut off parameter is due to the finite
lattice spacing $\rho_0 \sim \beta_1/\beta_2$ as well as to the contact
interaction. To find the exact value of $\rho_0$ one has to solve the
equation (\ref{17}) numerically with both short-range and long-range
interactions taken into account. Below we will present examples of such
calculations.

 Formula (\ref{En}) is one of the most important results in the paper.
It states that there are two (${\cal R}= \pm 1$) infinite series of bound
states for two holes with $S_z=0$. However let us remember that there are two
pockets in the Brillouin zone. Therefore we actually get four infinite sets of
large-size bound states. The dependence of binding energies on the hole
spin-wave
coupling constant $f$ and on the mass ratio $a$ manyfests itself in
Eq.(\ref{En})
through the parameter $\lambda$. For realistic value of $t\sim 3-4$ the
constant $F\approx 1$, see Fig.2. It makes both $\lambda_0$ and $\lambda_1$
not large: $|\lambda_0|\le 0.7$, $|\lambda_1|\le 1.5$, see Fig.4. Therefore
the bound states (\ref{En}) are shallow and long-range that.

Due to Eq.(\ref{En}) typical momenta in these states are small:
\begin{equation}
\label{q}
q\sim {1\over{\rho}}\sim {1\over{\rho}_0}
\exp\biggl(-{{\pi}\over{\sqrt{|\lambda|}}}(n+3/4)\biggr).
\end{equation}
Therefore the neglection of $(E-\epsilon_{\bf p}-\epsilon_{\bf k})$ in
comparison with $\omega_{\bf q}$ in Eq.(\ref{Gamma}) is justified.

\section{Symmetry properties}

  Let us discuss the relation connecting the parity ${\cal P}$ and the
quantum number ${\cal R}$. The wave function of relative motion in coordinate
representation is connected with that in momentum representation by
relation (\ref{Four}). The inversed relation is
\begin{equation}
\label{gk}
g_{\bf k} \propto \sum_n \psi({\bf r}_n)e^{-i{\bf k}{\bf r}_n}.
\end{equation}
However let us remember that
separation ${\bf r}_n$ between two holes of opposite spins is
always equal to the odd number of lattice spacings. The reason for this
is that the ground state of the system is an
antiferromagnetic one, see Eq.(\ref{9}).
To avoid misunderstanding let us recall that we are considering
dressed holes. Certainly due to the internal structure of the dressed
hole (\ref{10}) there are configurations with the separation between two bare
holes equal to the even number of lattice spacings. However ${\bf r}_n$
is the separation between the centers of dressed holes, which is always odd.
Thus only the terms with ${\bf r}_n$ corresponding to the odd number of
lattice spacings should be taken into account in sum (\ref{gk}).
Therefore
\begin{equation}
\label{gk1}
g_{{\bf k}-{\bf g}}=-g_{\bf k},
\end{equation}
where ${\bf g}=(\pm \pi,\pm \pi)$ is the inverse vector of the magnetic
lattice. Remember that the vector ${\bf k}$ is the deviation from ${\bf
k}_0$ which points to the band bottoms, ${\bf k}_0=(\pm \pi/2,
\pm\pi/2)$. The inversion is the transformation ${\bf k} \to -{\bf k},
{\bf k}_0 \to -{\bf k}_0 = {\bf k}_0-{\bf g}$. Therefore $g_{\bf k}$ is
transformed under the inversion as
\begin{equation}
\label{calP}
{\cal P}: g_{\bf k} \to g_{-{\bf k}-{\bf g}}
\end{equation}
Combining this with Eqs.(\ref{g}),(\ref{gk1}) we find
\begin{equation}
\label{calPR}
{\cal P}=-{\cal R}
\end{equation}
This is an important and surprising relation. Really, using the
definition of ${\cal R}$ (\ref{g}) it is easy to verify that it may be
presented as ${\cal R}=(-1)^l$, where $l$ is the number of nodes of the
angular wave function $\Phi(\varphi)$ in the region $0 \le \varphi \le
\pi$. Eq.(\ref{calPR}) states that the relation between $l$ and ${\cal
P}$ is unusual: ${\cal P}$ is positive (negative) for odd (even) $l$.

 It is useful to explane relation (\ref{calPR}) using coordinate
representation. The momentum with respect to the center of Brillouin
zone is equal ${\bf p}={\bf k}_0+{\bf k}$, where ${\bf k}$ is deviation
from the center of the pocket. Therefore
\begin{equation}
\label{psilarge}
\Psi({\bf r}_n)= \int g_{\bf k} e^{i{\bf p}{\bf r}_n}
{{d^2p}\over{(2\pi)^2}}=e^{i{\bf k}_0{\bf r}_n}
\int g_{\bf k} e^{i{\bf k}{\bf r}_n}
{{d^2k}\over{(2\pi)^2}}.
\end{equation}
Due to Eq.(\ref{Four}) the last integral here is $\psi({\bf r}_n)$, and
we get
\begin{equation}
\label{psilarge1}
\Psi({\bf r}_n)=\exp(i{\bf k}_0{\bf r}_n)\times \psi({\bf r}_n)=
\exp\biggl(i{{\pi}\over{2}}(n_x+n_y)\biggr)
\times \psi({\bf r}_n).
\end{equation}
Here ${\bf r}_n=(n_x,n_y)$. Since $n_x+n_y$ is odd the exponent
$\exp\biggl(i{{\pi}\over{2}}(n_x+n_y)\biggr)$  is equal to $\pm i$.
It changes the sign under reflection of coordinate
$n_x\to-n_x$, $n_y \to -n_y$.

 Relations (\ref{gk1}), (\ref{calPR}) permit one to find those
representations of the symmetry of the system in which the bound states
(\ref{En}) should manifest themselves. The group of symmetry of the
square lattice is ${\cal C}_{4v}$. It possesses four one-dimensional
representations $A_1, A_2, B_1, B_2$ of positive parity and one
two-dimensional representation $E$ of negative parity (see e.g.
Ref.\cite{Land}).

Consider first the bound state with $\lambda=\lambda_1$. It has ${\cal
R}=-1$ and hence the positive parity ${\cal P}=1$. Therefore, we are to
look for it among the one-dimensional representations. Further
restriction comes from (\ref{gk1}). In order to find it consider the
nodes of the wave function $g_{\bf k}$. We know that the angular
function $\Phi_{\lambda_1}(\varphi)$ has two nodes, $\varphi=0,\pi$,
which are at the axis perpendicular to the face of the Brillouin
zone, see Fig.5. The wave function in momentum representation $g_{\bf k}$
has a similar angular behaviour. It vanishes when $\bf k$
is orthogonal to the face of the zone. This property does not
contradict the symmetry in $A_2$ and $B_1$ representations, see Fig.6.
For $A_1$ and $B_2$ there is a trouble: the function $g_{\bf k}$ satisfying
(\ref{gk1}) in these representations must vanish at the edge, e.g. when
$\bf k$ is along the boundary. This means that the series of bound
states (\ref{En}) with $\lambda=\lambda_1$ have to appear
in $A_2$ and $B_1$ representations only.

A similar consideration shows that the bound states with
$\lambda=\lambda_0$ have to appear in two-dimensional $E$ representation.

We get the following full classification of the series of bound states
(\ref{En}): two of them with $\lambda=\lambda_1 ({\cal R}=-1, {\cal
P}=1)$ are in $B_1$ and $A_2$ representations, the other two with
$\lambda=\lambda_0 ({\cal R}=1, {\cal P}=-1)$ are degenerate, they
belong to $E$ representation.

\section{Numerical calculations}

  Now let us compare the analytical formula (\ref{En})
with the exact numerical solutions of Eq.(\ref{17}). Consider first the
states of positive parity with ${\cal R}=-1$. One series of them
belongs to the $B_1$ representation. The symmetry of the wave function
for this case is shown schematically in Fig.6a. The short-range
solution of this symmetry is usually called d-wave. The second
solution belongs to the $A_2$ representation (Fig.6b).

The numerical solution of Eq.(\ref{17}) is straightforward. We set the
parameter $\alpha$ in the contact hole-hole interaction
(\ref{Gamcon}),(\ref{12}):
$\alpha=0.6$; the mass ratio $a=\beta_1/\beta_2=7$. The value of hole-spin-wave
coupling constant is taken from Fig.2. The kernel $\Gamma$ is
calculated numerically with the help of Eqs.(\ref{Gam}), (\ref{Gamsw}),
(\ref{Gamcon}).

A few first energy levels of $B_1$ symmetry for different $t$
 are presented in Table 1. For each energy level we present also the
corresponding value of $\langle r^2 \rangle$
\begin{equation}
\label{r}
\langle r^2 \rangle = \int r^2 \psi^2({\bf r}) d^2r= \int |{{\partial
g_{\bf k}} \over {\partial {\bf k}}}|^2{{d^2k} \over{(2 \pi)^2}}.
\end{equation}
For $t=0$ there is only one short-range bound state:
$g_{\bf k}=\sqrt{2}(\cos k_x-\cos k_y)$.  It is due to
the contact hole-hole interaction.
In agreement with Refs.\cite{Ede2,Bon2,Poilblanc,Cher3} the short-range
bound state disappears at $t\approx 2$.

The energies of large-size states are given in the second and the third
sections of Table 1. The ratios $E_2/E_3$ are in agreement with formula
$E_n/E_{n+1}=\exp(2 \pi/\sqrt{|\lambda_1|})$ which follows from Eq.(\ref{En}).
Numerical values of $\exp(2 \pi/\sqrt{|\lambda_1|})$ with $\lambda_1$
determined by Figs.4,2 are presented in the last line of the Table 1.
The state at $t=2$ with $E=-0.0417$ is actually intermediate between short
range and long range one. Therefore the agreement in ratio $E_2/E_3$ is
not so good as for the other states.

 There are no short range bound states of $A_2$ symmetry. The situation
with long-range states is quite similar to that of $B_1$ symmetry:
they obey the Eq.(\ref{En}).

 The only difference for the states of negative parity (${\cal R}=+1$)
is that these states belong to the two-dimensional
representation $E$, therefore each state is doubly degenerate. At $t=0$
there is only short-range bound state $g_{\bf k}=2\sin k_x$ (or $g_{\bf
k}=2\sin k_y$), which is due to the contact hole-hole interaction. The
short-range state (usually it is called p-wave) disappears at $t\approx
2$ (see Refs.\cite{Ede2,Bon2,Poilblanc,Cher3}). At $t \ne 0$ there is the
series
of long-range states whose binding energies are described by Eq.(\ref{En}).

We conclude that there is a good agreement between the results of direct
numerical calculations and the analytical approach developed in
Sec.III.

\section{Conclusion}

In the present work the long-range ($l \gg lattice~ spacing$) positive and
negative parity two-hole bound states have been found for the holes on the
quantum N\'{e}el background. The parity of these states is related to the
angular wave function in a nonstandard way. Their small binding energy
depends exponentially on the hole-spin-wave coupling.

The natural question arises whether these states with the tiny binding
energy  could be relevant to high-$T_c$ superconductivity. There are
the strong physical reasons which show that it is quite possible. We
have demonstrated that there is the attraction between two holes caused
by the spin-wave exchange. We have considered these two holes
separately. If there is an ensemble of holes then one can expect that
the attraction between a pair of two holes should be enhanced. The
origin of this effect is the renormalization of spin-wave Green
function caused by the particle-hole excitations. Therefore the binding
energy of a pair is to be larger for the pair in the ensemble. The
point is that a quite moderate enhancement of the interaction results
in the very strong increase of the binding energy.

To illustrate the later statement let us consider $t=3$, and let us enhance the
hole-spin-wave coupling constant by a factor of $1.5$: $f=1.79 \to
f=2.69$. Then solving Eq.(\ref{17}) with this value of coupling
constant numerically we find the bound state of the pair belonging to
$B_1$ symmetry with the following characteristics:
\begin{equation}
\label{Co1}
E=-0.103,\hspace{2.1cm}\langle r_1^2 \rangle =8.4.
\end{equation}
This is the ground state with a reasonable energy of $E \sim 100 K$ (we
set $J=1 \equiv 0.13 eV$). The state is stretched along one direction,
and its maximal size is about 6-7 lattice spacings.

The physical picture of the interaction enhancement is to be discussed in
detail elsewhere.

{\bf ACKNOWLEDGMENTS}

  We are very grateful to V.V.Flambaum for valuable discussions. We
are also grateful very much to L.S.Kuchieva for the help in preparation of
the manuscript.

\begin{table}
\caption{The energies and the mean values of $r^2$ for a first few bound states
of $B_1$ symmetry}
\label{tab1}
\begin{tabular}{c|cccccc}
 t             & 0    &  1      &1.5     &  2     & 3      &  4     \\
\hline
$E_1$          &-0.250 & -0.262 &-0.146  &   -    &    -   &   -    \\
$\langle r_1^2 \rangle$ & 1   &1.22  & 1.95 & -   &    -   &   -    \\
\hline
$E_2$  & -  &$-1.84\cdot 10^{-3}$ & $-6.30\cdot 10^{-4}$ & $-4.17\cdot 10^{-2}$
 &$-2.21\cdot 10^{-4}$ & $-4.03\cdot 10^{-6}$\\
$\langle r_2^2 \rangle$ & -  &$1.74\cdot 10^2$ & $6.34\cdot 10^2$ & 7.83 &
$2.65 \cdot 10^3$ & $2\cdot 10^5$   \\
\hline
$E_3$  & -  &$-3.31\cdot 10^{-5}$ & $-5.25\cdot 10^{-6}$ & $-1.37\cdot 10^{-4}$
 &$\approx-2.\cdot 10^{-7}$ &  ? \\
$\langle r_3^2 \rangle$ & -  &$9.9\cdot 10^3$ & $8.5\cdot 10^4$ &
$3.5\cdot 10^3$ &
$\approx 2.8 \cdot 10^6$ & ?    \\
\hline
$(E_2/E_3)_{num}$ & - & 56 & 120 & 304 & 1105 & ? \\
$(E_n/E_{n+1})_{an}$ & - & 54 & 123 & 242 & 1071 & $2.15\cdot 10^4$ \\
\end{tabular}
\tablenotemark[1]{The first section presents the small-size bound state
which is due to the contact hole-hole interaction. It disappears at $t
\approx2$.}\\
\tablenotemark[2]{The large-size states are presented in the second and third
sections.}\\
\tablenotemark[3]{In the last section the ratio $(E_2/E_3)_{numerical}$ is
compared
with $(E_n/E_{n+1})_{analytical}$ predicted by formula (\ref{En}).}\\
\tablenotemark[4]{The accuracy of the code was insufficient to find
extremely shallow levels denoted by ``?''}
\end{table}

\newpage
{\bf FIGURE CAPTIONS}

FIG. 1. The Brillouin zone of a hole in the $t-J$ model.\\

FIG. 2. The coupling constants. Solid line represents the
hole-spin-wave coupling constant $f$ calculated in Ref.\cite{Suhf}.
Dashed line represents the effective constant $F=2f^2/(\pi \beta_1)$
which governs the long-range interaction between two holes, see Eq.(\ref{V}).\\

FIG. 3. Single spin-wave exchange between two holes. Its contribution to
the kernel of Eq.(\ref{17}) is given in (\ref{Gamsw}).\\

Fig. 4. The eigenvalues $\lambda$ of the angular equations (\ref{Phi}),
(\ref{bound}) versus $F$. The number at every curve is the value of the
mass ratio $a=\beta_1/\beta_2$=3,5,7,9. Fig.a - the solution $\lambda_0$
with ${\cal R}=+1$. Fig.b - the solution $\lambda_1$ with ${\cal R}=-1$.
Note that the vertical scale in Fig.b is different from that in Fig.a.\\

Fig. 5. The angular wave functions $\Phi(\varphi)$ for
$F=2f^2/(\pi \beta_1)=1.4$ and $a=\beta_1/\beta_2=7$.
Solid line is the solution whose eigenvalue is $\lambda_1$ with ${\cal
R}=-1$, and dashed line is the solution whose eigenvalue is $\lambda_0$
with ${\cal R}=1$.\\

Fig. 6. The symmetry of the positive parity (${\cal R}=-1$) bound state
wave function. Fig.a - $B_1$ type, Fig. b - $A_2$ type.
\end{document}